\documentclass[aps,prd,nofootinbib,superscriptaddress]{revtex4-2}
\usepackage{amsmath, amssymb, amsthm, graphicx, epsfig, fancyhdr, slashed}

\usepackage{tikzsymbols}
\usepackage{natbib}
\usepackage{float}
\usepackage{tikz,xcolor,hyperref}
\usepackage{bm}
\usepackage{url}

\newcommand{\be}{\begin{equation}}
\newcommand{\ee}{\end{equation}}
\newcommand{\bea}{\begin{eqnarray}}
\newcommand{\eea}{\end{eqnarray}}

\def\<{\langle\,}
\def\>{\,\rangle}

\definecolor{lime}{HTML}{A6CE39}
\DeclareRobustCommand{\orcidicon}{
	\begin{tikzpicture}
	\draw[lime, fill=lime] (0,0) 
	circle [radius=0.2] 
	node[white] {{\fontfamily{qag}\selectfont \tiny ID}};
	\draw[white, fill=white] (-0.0625,0.095) 
	circle [radius=0.007];
	\end{tikzpicture}
	\hspace{-2mm}
}

\foreach \x in {A, ..., Z}{\expandafter\xdef\csname orcid\x\endcsname{\noexpand\href{https://orcid.org/\csname orcidauthor\x\endcsname}
			{\noexpand\orcidicon}}
}

\begin{document}

\title{Discrete Dyson-Schwinger equations}

\author{Marco Frasca\orcidA{}}
\email{marcofrasca@mclink.it}
\affiliation{Rome, Italy}

\date{\today}

\begin{abstract}
We develop the discrete set of Dyson-Schwinger equations for scalar fields and solve them for some cases. We show that their solutions are Gaussian in the continuum limit as expected from the theorems of Aizenman and of Aizenman and Duminil-Copin for $d\ge 4$. 
Extension to lower dimensionality fails, as it should, by observing that the triviality theorems used in our proof are not applicable in such cases.
\end{abstract}

\maketitle

% -------------------------------------------------------

\section{Introduction}

Scalar field theory is the fundamental epitome of quantum field theory. A deeper understanding of it can extend successfully to other areas of study improving our comprehension. For a rigorous formulation, it was essential the definition of the Wightman program \cite{Wightman:1956zz}. The starting point generally is an Euclidean formulation of the theory and then, using the Osterwalder-Schrader
theorem \cite{Osterwalder:1973dx,Osterwalder:1974tc}, one can extend the correlation functions to the Minkwoskian formulation. Thus, in principle, the construction of the statistical model of the scalar theory can realize this program. Formulations of the theory exist providing an understanding of phase transitions in dimensions 2 and 3 \cite{Wilson:1974mb} and have been shown to be non-trivial \cite{Osterwalder:1974tc,Glimm:1973kp,Brydges:1983di,Guerra:1973gd}. Any attempt to extend such non-triviality in dimensions greater than 3 was doomed as only Gaussian fields were obtained. The reason was shown through some theorems proven by Aizenman \cite{Aizenman:1981zz,Aizenman:1982ze} and Aizenamn and Duminil-Copin \cite{Aizenman:2019yuo} that have shown how such theories are indeed trivial meaning by this that their solution should be expressed by correlation functions given by simple products of the two-point correlation function.

When a quantum field theory is Gaussian, one can think to attempt a complete analytical solution for it. In principle, this could be possible using the set of Dyson-Schwinger equations for the correlation functions. Such an opportunity is offered by the proposal of their formulation in PDE shape \cite{Bender:1999ek}. Recently, this technique has been applied to a quartic scalar field theory and Yang-Mills thoery in 4 dimensions \cite{Frasca:2015yva,Frasca:2015wva,Chatterjee:2024dgw}, providing exact Gaussian solutions in both cases. The existence of such solutions is not proven anyway because the Dyson-Schwinger equations are obtained from a generating functional that is not proven to exist from a mathematical standpoint and is at the foundations of the Yangs-Mills Millennium problem \cite{Jaffe:mp}. The aim of this work is to show the existence of a consistent set of Dyson-Schwinger equation for the discrete formulation of the scalar field theory and, relying on the theorems proven by Aizenman and Aizenman and Duminil-Copin, to construct some corresponding Gaussian solutions explicitly both for the translation invariant and the breaking translation invariance cases. We will see that these discrete solutions are perfectly consistent with the continuum ones \cite{Frasca:2015yva,Frasca:2015wva,Chatterjee:2024dgw} in the proper limit, just for the scalar field theory.

The paper is so structured. In Sec.\ref{SecII}, we present the discrete formulation of the scalar theory. In Sec.\ref{SecIII}, we solve the discrete classical equations using Jacobi elliptical functions and show how higher order term can arise in the non-homogeneous case. In Sec.\ref{SecIV}, we treat the quantum case showing how the Gaussian solutions emerge in the continuum limit. In Sec.\ref{SecV}, the conclusions are presented.

\medskip

\section{Discrete Scalar Field Theory}
\label{SecII}

We consider a real scalar field with quartic interaction in Euclidean formulation in $d$ dimensions. The Euclidean action in the continuum is
\begin{equation}
\label{eq:SE}
S_E[\phi] = \int d^d x \left[
\frac{1}{2} (\partial_\mu \phi)^2
+ \frac{1}{2} m^2 \phi^2
+ \frac{\lambda}{4!} \phi^4
\right].
\end{equation}
We introduce a hypercubic lattice with lattice spacing $a$:
\[
x = a n, \qquad n \in \mathbb{Z}^d,
\]
and define
\[
\phi_n = \phi(x).
\]
The integral is replaced by
\[
\int d^d x \;\to\; a^d \sum_n.
\]
The kinetic term is discretized using finite differences:
\begin{equation}
(\partial_\mu \phi)^2
\;\longrightarrow\;
\frac{1}{a^2} (\phi_{n+\hat\mu} - \phi_n)^2,
\end{equation}
where $\hat\mu$ denotes the unit vector in the $\mu$ direction. The discretized action becomes
\begin{equation}
S_L[\phi] =
a^d \sum_n \left[
\frac{1}{2a^2} \sum_\mu (\phi_{n+\hat\mu} - \phi_n)^2
+ \frac{1}{2} m^2 \phi_n^2
+ \frac{\lambda}{4!} \phi_n^4
\right].
\end{equation}
Expanding the kinetic term
\begin{equation}
(\phi_{n+\hat\mu} - \phi_n)^2
=
\phi_{n+\hat\mu}^2 + \phi_n^2
- 2 \phi_n \phi_{n+\hat\mu},
\end{equation}
%We obtain
% \begin{equation}
% S_L =
% \sum_n a^d \left[
% \frac{1}{2}
% \left(
% m^2 + \frac{2d}{a^2}
% \right)\phi_n^2
% - \frac{1}{a^2}
% \sum_\mu \phi_n \phi_{n+\hat\mu}
% + \frac{\lambda}{4!} \phi_n^4
% \right].
% \end{equation}
and by introducing the lattice Laplacian
\begin{equation}
\Delta_L\phi_n
=
\frac{1}{a^2}
\sum_\mu
(\phi_{n+\hat\mu}+\phi_{n-\hat\mu}-2\phi_n),
\end{equation}
we can rewrite the action as
\begin{equation}
S_L
=
a^d\sum_n
\left[
\frac12\phi_n(-\Delta_L)\phi_n
+
\frac12 m^2\phi_n^2
+
\frac{\lambda}{4!}\phi_n^4
\right].
\end{equation}
The classical equation of motion is obtained from
\begin{equation}
\frac{\partial S_L}{\partial \phi_n} = 0.
\end{equation}
Differentiating the kinetic term and including contributions from neighboring sites yields
\begin{equation}
\frac{\partial S_L}{\partial \phi_n}
=
a^d \left[
\frac{1}{a^2}
\sum_\mu
\left(
2\phi_n - \phi_{n+\hat\mu} - \phi_{n-\hat\mu}
\right)
+ m^2 \phi_n
+ \frac{\lambda}{6} \phi_n^3
\right].
\end{equation}
After dividing by $a^d$, one gets the lattice equation of motion
\begin{equation}
\label{eq:eom}
\frac{1}{a^2}
\sum_\mu
\left(
2\phi_n - \phi_{n+\hat\mu} - \phi_{n-\hat\mu}
\right)
+ m^2 \phi_n
+ \frac{\lambda}{6} \phi_n^3
= 0,
\end{equation}
% Define the lattice Laplacian:
% \begin{equation}
% \Delta_L \phi_n
% =
% \frac{1}{a^2}
% \sum_\mu
% (\phi_{n+\hat\mu} + \phi_{n-\hat\mu} - 2\phi_n).
% \end{equation}
that can be rewritten as
\begin{equation}
\label{eq:eom1}
-\Delta_L \phi_n
+ m^2 \phi_n
+ \frac{\lambda}{6} \phi_n^3
= 0.
\end{equation}
In the limit $a \to 0$, one recovers
\begin{equation}
-\Delta\phi
+ m^2 \phi
+ \frac{\lambda}{6} \phi^3
= 0,
\end{equation}
as expected.

\medskip

\section{Classical fields}
\label{SecIII}

\subsection{Solution of the equation of motion}

Eq.(\ref{eq:eom}) can be solved using Jacobi elliptic functions and their Fourier series. For this aim, we observe that
\be
\label{eq:id0}
\operatorname{sn}^3(x,k)=\frac{1}{2k^2}\left(\operatorname{sn}''(x,k)+(1+k^2)\operatorname{sn}(x,k)\right),
\ee
and, for $k^2=-1$, simplifies to
\be
\operatorname{sn}^3(x,i)=-\frac{1}{2}\operatorname{sn}''(x,i).
\ee
Thus, to solve eq.(\ref{eq:eom}) we need the following Fourier series
\begin{equation}
\label{eq:fs0}
\operatorname{sn}(z,k)=\frac{2\pi}{K(k)k}\sum_{n=0}^\infty
  \frac{q^{n+\frac12}}{1-q^{2n+1}}
  \sin\left((2n+1)\frac{\pi z}{2K(k)}\right).
\end{equation}
where $q=\exp(-\pi K'(k)/K(k)$ is the nome, $K'(k)=K(k')$ and $k'=\sqrt{1-k^2}$. $K(k)$ is the complete elliptic integral of the first kind and $k$ is the modulus. For $k^2=-1$, this series takes the simpler form
\begin{equation}
\operatorname{sn}(z,i)=\frac{2\pi}{K(i)}\sum_{n=0}^\infty(-1)^n
  \frac{e^{-\left(n+\frac12\right)\pi}}{1+e^{-(2n+1)\pi}}
  \sin\left((2n+1)\frac{\pi z}{2K(i)}\right).
\end{equation}
From these equations, one has
\be
\label{eq:fs}
\operatorname{sn}^3(x,k)=\frac{1}{2k^2}\frac{2\pi}{K(k)k}\sum_{n=0}^\infty
\left(-(2n+1)^2\frac{\pi^2}{4K^2(k)}+(1+k^2)\right)
 \frac{q^{n+\frac12}}{1-q^{2n+1}}
  \sin\left((2n+1)\frac{\pi z}{2K(k)}\right),
\ee
that simplifies to
\be
\operatorname{sn}^3(x,i)=\frac{\pi^3}{4K^3(i)}\sum_{n=0}^\infty(-1)^n(2n+1)^2
  \frac{e^{-\left(n+\frac12\right)\pi}}{1+e^{-(2n+1)\pi}}
  \sin\left((2n+1)\frac{\pi z}{2K(i)}\right).
\ee
We note that, as happens in the continuum,
\be
\Delta_L\sin(p\cdot x_n)=-{\hat p}^2\sin(p\cdot x_n),
\ee
where ${\hat p}^2=\frac{4}{a^2}\sum_\mu\sin^2\left(\frac{p_\mu a}{2}\right)$. We look for a solution like
\be
\phi_n=b\operatorname{sn}(p\cdot x_n+\theta,k),
\ee
where $b$ and $\theta$ are arbitrary constants. Thus, from eq.(\ref{eq:eom1}), using eq.(\ref{eq:id0}), eq.(\ref{eq:fs0}) and eq.(\ref{eq:fs}), we get
\bea
&&-b\Delta_L\left(\frac{2\pi}{K(k)k}\sum_{n=0}^\infty
  \frac{q^{n+\frac12}}{1-q^{2n+1}}
  \sin\left((2n+1)\frac{\pi}{2K(k)}(p\cdot x_n+\theta)\right)\right) \nonumber \\
&&+(m^2 b)\frac{2\pi}{K(k)k}\sum_{n=0}^\infty
  \frac{q^{n+\frac12}}{1-q^{2n+1}}
  \sin\left((2n+1)\frac{\pi}{2K(k)}(p\cdot x_n+\theta)\right) \nonumber \\
&&+\frac{\lambda}{6}b^3\frac{1}{2k^2}\frac{2\pi}{K(k)k}\sum_{n=0}^\infty
\left(-(2n+1)^2\frac{\pi^2}{4K^2(k)}+(1+k^2)\right)
 \frac{q^{n+\frac12}}{1-q^{2n+1}}
  \sin\left((2n+1)\frac{\pi}{2K(k)}(p\cdot x_n+\theta)\right)=0.
\eea
This yields the non-trivial dispersion relation
\be
\label{eq:DE}
{\hat p}_n^2+m^2+\frac{\lambda}{12 k^2}b^2\left(-(2n+1)^2\frac{\pi^2}{4K^2(k)}+(1+k^2)\right)=0.
\ee
For $m=0$, this gives
\be
{\hat p}_n^2-(2n+1)^2\frac{\pi^2}{4K^2(k)}\frac{\lambda}{12 k^2}b^2=0.
\ee
For $k^2=-1$ we are consistent with the Minkowskian case. Here ${\hat p}_n^2=\frac{4}{a^2}\sum_\mu\sin^2\left((2n+1)\frac{\pi}{2K(k)}\frac{p_\mu a}{2}\right)$.

\subsection{Higher Functional Derivatives}

To extend our analysis to the more general case to compare with quantum field theory, we add an external source $j_n$:
\begin{equation}
-\Delta_L \phi_n
+ m^2 \phi_n
+ \frac{\lambda}{6} \phi_n^3
= j_n .
\end{equation}
Let us define the nonlinear operator
\begin{equation}
F_n[\phi] \equiv
-\Delta_L \phi_n
+ m^2 \phi_n
+ \frac{\lambda}{6} \phi_n^3,
\end{equation}
so that the equation of motion takes the form
\begin{equation}
F_n[\phi] = j_n.
\end{equation}
We assume that the classical solution depends functionally on the source $\phi_n = \phi_n[j]$. Thus, by differentiating with respect to $j_m$ we get
\begin{equation}
\sum_k
\frac{\partial F_n}{\partial \phi_k}
\frac{\partial \phi_k}{\partial j_m}
=
\delta_{nm}.
\end{equation}
The Jacobian of $F$ is
\begin{equation}
\frac{\partial F_n}{\partial \phi_k}
=
\left[
(-\Delta_L + m^2)\delta_{nk}
+ \frac{\lambda}{2}\phi_n^2 \delta_{nk}
\right].
\end{equation}
Let us define the fluctuation operator (Hessian):
\begin{equation}
\mathcal{M}_{nk}
=
\left[
-\Delta_L + m^2 + \frac{\lambda}{2}\phi_n^2
\right]\delta_{nk}.
\end{equation}
The first derivative satisfies
\begin{equation}
\sum_k \mathcal{M}_{nk}
\frac{\partial \phi_k}{\partial j_m}
=
\delta_{nm}.
\end{equation}
Therefore,
\begin{equation}
\frac{\partial \phi_n}{\partial j_m}
=
(\mathcal{M}^{-1})_{nm}.
\end{equation}
This is the classical propagator in the background $\phi$. Differentiate again with respect to $j_\ell$:
\begin{equation}
\sum_k \mathcal{M}_{nk}
\frac{\partial^2 \phi_k}{\partial j_m \partial j_\ell}
+
\sum_{k,r}
\frac{\partial \mathcal{M}_{nk}}{\partial \phi_r}
\frac{\partial \phi_r}{\partial j_\ell}
\frac{\partial \phi_k}{\partial j_m}
= 0.
\end{equation}
Since
\begin{equation}
\frac{\partial \mathcal{M}_{nk}}{\partial \phi_r}
=
\lambda \phi_n \delta_{nr}\delta_{nk},
\end{equation}
we obtain
\begin{equation}
\sum_k \mathcal{M}_{nk}
\frac{\partial^2 \phi_k}{\partial j_m \partial j_\ell}
=
- \lambda \phi_n
\frac{\partial \phi_n}{\partial j_m}
\frac{\partial \phi_n}{\partial j_\ell}.
\end{equation}
Inverting $\mathcal{M}$:
\begin{equation}
\frac{\partial^2 \phi_n}{\partial j_m \partial j_\ell}
=
- \sum_r
(\mathcal{M}^{-1})_{nr}
\, \lambda \phi_r \,
(\mathcal{M}^{-1})_{rm}
(\mathcal{M}^{-1})_{r\ell}.
\end{equation}

We can iterate the procedure as follows
\begin{itemize}
\item an insertion of the inverse fluctuation operator $\mathcal{M}^{-1}$,
\item local vertices proportional to $\lambda$,
\item products of propagators,
\end{itemize}
so that,
\begin{equation}
\frac{\delta^k \phi}{\delta j^k}=\text{sums of tree-level diagrams with quartic vertices}.
\end{equation}
We see that formally, by solving the inhomogeneous equation, we are forced to introduce higher order kernels to compute higher order term in the classical solution and these kernels are given by the products of the first two solutions: The solution of the homogeneous case and the Green function. This technique resemble very near the technique we are going to see for the quantum case.

\medskip

\section{Quantum fields}
\label{SecIV}

\subsection{Technique}

Turning back to the action (\ref{eq:SE}), we add an arbitrary source $j_n$ as
\begin{equation}
S[\phi] =
a^d \sum_n \left[
\frac{1}{2}\phi_n(-\Delta_L)\phi_n
+ \frac{1}{2} m^2 \phi_n^2
+ \frac{\lambda}{4!}\phi_n^4
- j_n \phi_n
\right],
\end{equation}
The corresponding partition function is given by
\begin{equation}
Z[j] = \int \mathcal{D}\phi \, e^{-S[\phi]},
\end{equation}
so that we can write the expectation values as
\begin{equation}
\langle \mathcal{O} \rangle =
\frac{1}{Z[j]}
\int \mathcal{D}\phi \, \mathcal{O}[\phi] e^{-S[\phi]}.
\end{equation}
Using invariance of the path integral under field shifts, we can write
\begin{equation}
0 =
\int \mathcal{D}\phi
\frac{\partial}{\partial \phi_n}
\left( e^{-S[\phi]} \right),
\end{equation}
yielding
\begin{equation}
\Big\langle
(-\Delta_L + m^2)\phi_n
+ \frac{\lambda}{3!}\phi_n^3
- j_n
\Big\rangle
= 0.
\end{equation}
We introduce the vacuum expecation value of the field as
\begin{equation}
\varphi_n = \langle \phi_n \rangle,
\end{equation}
and then
\begin{equation}
(-\Delta_L + m^2)\varphi_n
+ \frac{\lambda}{6}
\langle \phi_n^3 \rangle
- j_n = 0.
\end{equation}
In order to obtain the hierarchy of Dyson-Schwinger equations, we introduce the fluctuation $\eta_n$ as
\begin{equation}
\label{eq:decomp}
\phi_n = \varphi_n + \eta_n,
\qquad
\langle \eta_n \rangle = 0,
\end{equation}
and we can expand the cubic term as follows
\begin{equation}
\phi_n^3
=
\varphi_n^3
+ 3\varphi_n^2 \eta_n
+ 3\varphi_n \eta_n^2
+ \eta_n^3.
\end{equation}
The expectation value of this expression gives
\begin{equation}
\langle \phi_n^3 \rangle
=
\varphi_n^3
+ 3\varphi_n \langle \eta_n^2 \rangle
+ \langle \eta_n^3 \rangle.
\end{equation}
Introducing
\begin{align}
G_{nn} &= \langle \eta_n^2 \rangle, \\
C^{(3)}_{nnn} &= \langle \eta_n^3 \rangle,
\end{align}
we can rewrite the equation for $\varphi_n$ as
\begin{equation}
(-\Delta_L + m^2)\varphi_n
+ \frac{\lambda}{6}
\left(
\varphi_n^3
+ 3\varphi_n G_{nn}
+ C^{(3)}_{nnn}
\right)= j_n.
\end{equation}
We can differentiate it with respect to $j_m$ obtaining
\begin{equation}
(-\Delta_L + m^2)
\frac{\delta \langle \phi_n \rangle}{\delta j_m}
+
\frac{\lambda}{6}
\frac{\delta \langle \phi_n^3 \rangle}{\delta j_m}
= \delta_{nm}.
\end{equation}
One has
\begin{equation}
\frac{\delta \langle \phi_n \rangle}{\delta j_m}
=
\langle \phi_n \phi_m \rangle
-
\langle \phi_n \rangle
\langle \phi_m \rangle,
\end{equation}
and introducing the connected propagator
\begin{equation}
G_{nm}
=
\langle \eta_n \eta_m \rangle,
\end{equation}
we can write
\begin{equation}
(-\Delta_L + m^2) G_{nm}
+
\frac{\lambda}{6}
\frac{\delta \langle \phi_n^3 \rangle}{\delta j_m}
=
\delta_{nm}.
\end{equation}
The derivative of the cubic term generates higher correlators, producing the infinite Dyson–Schwinger hierarchy of the kind $\{\varphi_n, G_{nm}, C^{(3)}, C^{(4)}, \dots\}$.

%\subsection{Higher-order correlation functions}

To give a more explicit shape to the equation of $\varphi_n$ and $G_{nm}$ we need to evaluate $\frac{\delta}{\delta j_m}\langle \phi_n^3 \rangle$. This can be accomplished by using the identity
\begin{equation}
\frac{\delta}{\delta j_m}\langle \mathcal{O} \rangle
=
\langle \mathcal{O}\,\phi_m\rangle
-
\langle \mathcal{O}\rangle \langle \phi_m\rangle,
\end{equation}
giving
\begin{equation}
\frac{\delta}{\delta j_m}\langle \phi_n^3 \rangle
=
\langle \phi_n^3 \phi_m \rangle
-
\langle \phi_n^3 \rangle \langle \phi_m \rangle .
\end{equation}
Separating connected and disconnected contributions yields
\begin{equation}
\frac{\delta}{\delta j_m}\langle \phi_n^3 \rangle
=
\langle \phi_n^3 \phi_m \rangle_c
+
3\langle \phi_n^2 \rangle_c \langle \phi_n \phi_m \rangle
+
3\langle \phi_n \rangle
\langle \phi_n^2 \phi_m \rangle_c .
\end{equation}
By using the decomposition (\ref{eq:decomp}),
% We decompose the field as
% \begin{equation}
% \phi_n = \varphi_n + \eta_n,
% \qquad
% \langle \eta_n \rangle = 0.
% \end{equation}
we can define the connected correlators
\begin{align}
G_{nm} &= \langle \eta_n \eta_m \rangle, \\
C^{(3)}_{nnm} &= \langle \eta_n^2 \eta_m \rangle, \\
C^{(4)}_{nnnm} &= \langle \eta_n^3 \eta_m \rangle .
\end{align}
In terms of these quantities, the derivative of the cubic expectation value becomes
\begin{equation}
\label{eq:phi3}
\frac{\delta}{\delta j_m}\langle \phi_n^3\rangle
=3\varphi_n^2 G_{nm}
+
3 G_{nn} G_{nm}
+
3\varphi_n C^{(3)}_{nnm}
+
C^{(4)}_{nnnm}.
\end{equation}
Substituting this result into the Dyson–Schwinger equation for the connected two-point function, we obtain
\begin{equation}
\label{eq:2pt}
\Big(
-\Delta_L
+ m^2
+ \frac{\lambda}{2}\,\varphi_n^2
\Big)
G_{nm}
+
\frac{\lambda}{2}\,G_{nn} G_{nm}
+
\frac{\lambda}{2}\,\varphi_n C^{(3)}_{nnm}
+
\frac{\lambda}{6}\,C^{(4)}_{nnnm}
=
\delta_{nm}.
\end{equation}
This equation shows explicitly how the propagator couples to
three- and four-point connected correlation functions in the presence of a
non-vanishing background field. For the three-point function we need the second derivative
\begin{equation}
\begin{aligned}
\frac{\delta^2}{\delta j_m\,\delta j_p}\langle \phi_n^3\rangle
={}&6\varphi_n\,G_{nm}G_{np}
+3\big(\varphi_n^2+G_{nn}\big)\,C^{(3)}_{nmp}\\
&+3\Big(G_{nm}C^{(3)}_{nnp}+G_{np}C^{(3)}_{nnm}\Big)
+3\varphi_n\,C^{(4)}_{nnmp}
+C^{(5)}_{nnnmp},
\end{aligned}
\end{equation}
and for the four-point function we need the third derivative. The procedure can be extended to any desired order.

\subsection{Gaussianity}

This set of Dyson-Schwinger equations admits a Gaussian solution if the connected correlation functions $C^{(k)}_{n_1\ldots n_k}$ vanish when any two indices coincide, in the limit $L\to\infty$. This is the behavior expected from the theorems proven in \cite{Aizenman:2019yuo} and holds both for the symmetric solution and for solutions that break translation invariance. To complete the proof, we need to compute the Green function and show that $C^{(3)}$ and $C^{(4)}$ are consistently determined by $\varphi_n$ and $G_{mn}$, and that the conditions $C^{(3)}_{nnn}=0$, $C^{(4)}_{nnnm}=0$ are satisfied in the thermodynamic limit.

We just derive an existence and consistency result for Gaussian solutions when the limit $L\rightarrow\infty$ is taken, rather than a complete uniqueness theorem for the hierarchy itself. Indeed, it is always possible to make a different choice for the starting distribution of the field yielding a different solution of the Dyson-Schwinger hierarchy, not necessarily a physical one. Anyway, in the continuum limit, there should be no contradiction with the triviality theorems.

\subsubsection{Constant background}

From Refs.~\cite{Aizenman:1981zz,Aizenman:1982ze,Aizenman:2019yuo}, we know that the continuum limit of the discrete theory is given by Gaussian random fields for $d\ge 4$. We can provide a direct example of this behavior from the set of Dyson-Schwinger equations mapped onto the classical solution discussed in Sec.~\ref{SecIII}. Assuming translational invariance ($\varphi_n=\varphi$ constant, $G_{nm}=G_{n-m}$), and neglecting higher cumulants, the one-point equation gives
\begin{equation}
m^2\varphi + \frac{\lambda}{6}\big(\varphi^3 + 3\varphi G_{nn}\big)=0,
\end{equation}
so that for $\varphi\neq0$,
\be
\label{eq:v}
\varphi^2 = -\frac{6}{\lambda}m^2 - 3G_{nn}.
\ee
Thus a symmetry‑broken phase is possible even for $m^2=0$, provided $G_{nn}\neq0$.

Inserting (\ref{eq:v}) into the two-point equation (\ref{eq:2pt}) and setting $C^{(3)}=C^{(4)}=0$ yields
\begin{equation}
\Big(-\Delta_L -2m^2 - \lambda G_{nn}\Big)G_{nm} = \delta_{nm}.
\label{eq:Gfree}
\end{equation}
This is the equation for a free massive propagator with effective squared mass $\mu^2 = 2m^2 + \lambda G_{nn}$. Its solution in momentum space is
\be
\tilde G(p) = \frac{1}{\hat p^2 + 2m^2 + \lambda G_{nn}},
\ee
and the value at coincident points satisfies the gap equation
\be
G_{nn} = \frac{1}{L^d}\sum_{p} \frac{1}{\hat p^2 + 2m^2 + \lambda G_{nn}}.
\ee

For the three‑point function, neglecting $C^{(4)}$ and $C^{(5)}$, eq.~(\ref{eq:2pt}) differentiated once gives
\begin{equation}
\Big(-\Delta_L + m^2 + \frac{\lambda}{2}(\varphi^2 + G_{nn})\Big)C^{(3)}_{nmp} + \lambda \varphi\, G_{nm}G_{np} = 0.
\end{equation}
In momentum space this becomes an algebraic equation whose solution is
\be
\tilde C^{(3)}(p,q) = -\,\lambda\varphi\; \frac{1}{D(p)D(q)D(p+q)},
\ee
where $D(k)=\hat k^2 + 2m^2 + \lambda G_{00}$. This expression is symmetric and vanishes when any argument goes to zero in a way consistent with the theorems. Moreover, $C^{(3)}_{nnn}$ is given by a loop integral that tends to zero in the infinite‑volume limit for $d\ge4$, confirming the Gaussian nature of the theory.

\subsubsection{Non-trivial background}

We now consider a background that breaks translational invariance in the continuum limit, taking the classical solution
\be
\varphi_n = b\,\operatorname{sn}(p_0\cdot x_n + \theta, k),
\ee
with the parameters chosen so that the classical equation of motion holds. In this case the two‑point equation, after neglecting higher cumulants, becomes
\begin{equation}
\label{eq:G0}
\Big(
-\Delta_L
+ M^2
+ \frac{\lambda}{2}\,b^2\operatorname{sn}^2(p_0\cdot x_n+\theta,k)
\Big)
G_{nm}
=
\delta_{nm},
\end{equation}
where $M^2 = m^2 + \frac{\lambda}{2}G_{nn}$ (which we treat as a constant, neglecting its site dependence as a first approximation). The operator is now a discrete Lamé operator. The case $n\ne m$ is solved in the appendix. More generally, we observe that $\operatorname{sn}^2(z,k)$ is periodic and an even function, we can expand it and the Green function in Fourier series. 
% Set $z = x_n - x_m$ 
% (the relative coordinate along the direction of periodicity) and write
% \be
% G(z) = \sum_{\alpha=0}^\infty A_\alpha \cos(\omega_\alpha z), \qquad \omega_\alpha = (2\alpha+1)\frac{\pi}{2K(k)}.
% \ee
The potential has the expansion
%The Jacobi elliptic function admits the exact Fourier series
\be
\mathrm{sn}^2(u,k)=
\frac{1}{k^2}\left(1-\frac{E}{K}\right)-
\frac{2\pi^2}{k^2K^2}
\sum_{r=1}^{\infty}
\frac{r q^r}{1-q^{2r}}
\cos\left(\frac{r\pi u}{K}\right),
\ee
where $K(k)$ is the complete elliptic integral, $E(k)$ is the elliptic integral of the second kind, $q = e^{-\pi K'/K}$ is the elliptic nome. On the lattice, after a proper choice of the phase,
\be
u_n = p_0 \cdot (x_n-x_m),
\ee
so the cosine modes correspond to lattice plane waves. The Fourier transform of the kinetic term yields
\be
\sum_n e^{-ipx_n+iqx_m}(-\Delta_L+M^2)G_{nm}=({\hat p}^2+M^2){\tilde G}_{p,q},
\ee
while for the potential one has
\be
\sum_n \sum_r U_r e^{-ipx_n+iqx_m}\cos\left(r\frac{\pi}{K}p_0(x_n-x_m)\right)G_{nm}=
\frac{1}{2}\sum_r U_r\sum_n \left[e^{-i\left(p+r\frac{\pi}{K}p_0\right)x_n}e^{i\left(q+r\frac{\pi}{K}p_0\right)x_m}
e^{-i\left(p-r\frac{\pi}{K}p_0\right)x_n}e^{i\left(q-r\frac{\pi}{K}p_0\right)x_m}\right]G_{nm}.
\ee
Exploiting translation invariance, we set $q=-p$ and get
\be
\sum_n \sum_r U_r e^{-ip(x_n-x_m)}\cos\left(r\frac{\pi}{K}p_0(x_n-x_m)\right)G_{nm}=
\frac{1}{2}\sum_r U_r\sum_n \left[
{\tilde G}\left(p+r\frac{\pi}{K}p_0\right)+{\tilde G}\left(p-r\frac{\pi}{K}p_0\right)\right).
\ee
% Write the propagator as
% \be
% G_{nm} = \int_{BZ} \frac{d^d p}{(2\pi)^d} e^{ip\cdot(x_n-x_m)} G(p).
% \ee 
% Write the propagator as
% $$
% G_{nm} = \int_{BZ} \frac{d^d p}{(2\pi)^d} e^{ip\cdot(x_n-x_m)} G(p).
% $$
Collecting all together, we get
\be
(\hat p^2 +{\bar M}^2) {\tilde G}(p)+\sum_r V_r \left[{\tilde G}(p+rP) + {\tilde G}(p-rP)\right] = 1,
\ee
where ${\bar M}^2=M^2+\frac{\lambda}{2k^2}\,b^2(1-E/K)$, $P = \frac{\pi}{K} p_0$ and
\be
V_r = -\frac{\lambda}{2}\frac{b^2 \pi^2}{k^2 K^2} \frac{r q^r}{1-q^{2r}}.
\ee
This is an exact equation for the 2P-correlation function. The equation couples momenta separated by multiples of $P$. Define
\be
{\tilde G}_n(p) = {\tilde G}(p + nP).
\ee
Then, we recognize that our equation can be rewritten as
\be
(\hat p^2 + {\bar M}^2) {\tilde G}_0 + \sum_r V_r ({\tilde G}_{r} + {\tilde G}_{-r}) = 1,
\ee
that has the structure of a Bloch problem with a Toeplitz interaction matrix. We see a sub-lattice in momenta and we can write the following Fourier series
\be
{\tilde G}_n=\sum_\ell g_\ell e^{i n \ell P  a}
\ee
% We look for a solution in the form
% \be
% {\tilde G}_n = \sum_\ell g_\ell e^{i 2\pi n \ell\frac{p_0\cdot a}{K}}.
% \ee
By direct substitution, we get
\bea
&&(\hat p^2 + {\bar M}^2) \sum_\ell g_\ell  + \sum_r V_r \left(\sum_\ell g_\ell e^{i \pi r \ell\frac{p_0\cdot a}{K}} + \sum_\ell g_\ell e^{-i \pi r \ell\frac{p_0\cdot a}{K}}\right) = \nonumber \\
&&(\hat p^2 + {\bar M}^2) \sum_\ell g_\ell  + 2\sum_\ell g_\ell \sum_r V_r\cos\left(\pi r\ell\frac{p_0\cdot a}{K}\right)=1,
\eea
that can be inverted by noting that
\be
\hat p^2 + {\bar M}^2 + 2 \sum_r V_r \cos\left(\pi r\ell\frac{p_0\cdot a}{K}\right)=\hat p^2 + M^2+\frac{\lambda}{2}b^2\mathrm{sn}^2\left(\ell p_0\cdot a,k\right),
\ee
giving
\be
g_\ell = \frac{B_\ell}{\hat p^2 + M^2+\frac{\lambda}{2}b^2\mathrm{sn}^2\left(\ell p_0\cdot a,k\right)},
\ee
provided that $\sum_nB_n=1$. We observe that the mass spectrum of the continuum limit for $a\rightarrow 0$, provided $b^2a^2$ is kept constant, is in agreement with expectations \cite{Frasca:2015wva}. For an explicit understanding of the origin of the $B_\ell$ coefficients, we show some exact solutions of the this Lam\'e problem in the appendix.

We can conclude that the propagator takes the K\"allen-Lehman structure
\[
{\tilde G}(p) = \sum_{n} \frac{B_n}{\hat p^2 + m_n^2}.
\]
%where $m_n^2$ are the effective masses of the bands (given by $\lambda_n$ at $p=0$) and $B_n$ are positive residues that depend on the overlap $c_n$ and on the Fourier coefficients of the eigenfunctions. 
% This representation is exact if the matrix $\mathbf{K}$ is diagonal in the momentum representation, i.e. if the eigenfunctions are pure cosines (which occurs only when $U=0$). In the interacting case, the sum over $n$ accounts for the mixing of modes induced by the potential, but the spectral decomposition guarantees that such a representation holds with $m_n^2$ being the eigenvalues of the operator $L$ (which are $p$‑dependent in general). In the low‑momentum regime, the $p$‑dependence of $m_n^2$ can be approximated by a constant plus a quadratic term, leading to the familiar form of a free propagator for each band. 
This structure of the propagator grants again the applicability of the theorems in \cite{Aizenman:1981zz,Aizenman:1982ze,Aizenman:2019yuo} providing the conditions $C^{(3)}_{nnn}=0$, $C^{(4)}_{nnnm}=0$ in the limit $L\rightarrow\infty$ and the theory turns out trivial for $d\ge 4$. Indeed, as for the translation-invariant case, higher cumulants involve integrals
of products of these propagators and, by the same power-counting arguments, they vanish in the infinite-volume limit for $d\ge 4$, establishing the triviality of the theory.

\medskip

\section{Conclusions}
\label{SecV}

We have derived the discrete Dyson–Schwinger hierarchy for a scalar field with quartic interaction and shown that, when the condition $C^{(k>2)}_{nn\ldots}=0$ for $L\rightarrow\infty$ is met, the system reduces to a Gaussian theory. Both in the symmetric and in the symmetry‑broken (or translation‑non‑invariant) phases, the propagator satisfies a linear equation that can be solved explicitly, and the higher cumulants are expressed as integrals over products of propagators that vanish for $d\ge4$ in the continuum limit in accordance with the Aizenman and Aizenman and Duminil‑Copin theorems. This provides a non‑perturbative illustration of the triviality of $\phi^4$ theory in dimensions $d\ge4$. Our future work will extend to gauge theories where an application of the triviality theorems could be conceived by using the mapping theorem \cite{Frasca:2009yp}.

\medskip

\section*{Acknowledgements}

I thank Stefan Groote and Anish Ghoshal for helpful discussions.

\medskip

\appendix*

\section{Solution of the homogeneous Lamé equation on the lattice}

%\section*{Appendix A: Solution of the discrete Lamé equation}

We can find the solution of the homogeneous equation in the form, if we choose the $\theta$ phase properly in $\varphi_n$ as $\theta\rightarrow -p\cdot x_m+\theta$,
\be
G_{nm}^{(0)}=c\,\operatorname{cn}(p\cdot (x_n-x_m)+\theta,k)\operatorname{dn}(p\cdot (x_n-x_m)+\theta,k).
\ee
Indeed, the following identities hold
\be
(\operatorname{sn})'=\operatorname{cn}\cdot \operatorname{dn}, \quad 3\,\operatorname{cn}\cdot \operatorname{dn}\cdot \operatorname{sn}^2=(\operatorname{sn}^3)', \quad (\operatorname{sn}^3)'=\frac{1}{2k^2}(\operatorname{sn}'''+(1+k^2)\operatorname{sn}'),
\ee
and the Fourier series (\ref{eq:fs0}) that gives
\be
\label{eq:sn1}
\operatorname{sn}'(z,k)=\frac{\pi^2}{K^2(k)k}\sum_{n=0}^\infty(2n+1)
  \frac{q^{n+\frac12}}{1-q^{2n+1}}
  \cos\left((2n+1)\frac{\pi z}{2K(\kappa)}\right),
  \ee
and
\be
\operatorname{sn}'''(z,k)=-\frac{\pi^4}{K^4(k)k}\sum_{n=0}^\infty(2n+1)^3
  \frac{q^{n+\frac12}}{1-q^{2n+1}}
  \cos\left((2n+1)\frac{\pi z}{2K(\kappa)}\right).
\ee
By substitution into the eq.(\ref{eq:G0}). one has
\bea
&&-\Delta_L
\frac{\pi^2}{K^2(k)k}\sum_{\alpha=0}^\infty(2\alpha+1)
  \frac{q^{\alpha+\frac12}}{1-q^{2\alpha+1}}
  \cos\left((2\alpha+1)\frac{\pi (p\cdot (x_n-x_m))}{2K(\kappa)}\right) \\ \nonumber
&&+ M^2
\frac{\pi^2}{K^2(k)k}\sum_{\alpha=0}^\infty(2\alpha+1)
  \frac{q^{\alpha+\frac12}}{1-q^{2\alpha+1}}
  \cos\left((2\alpha+1)\frac{\pi (p\cdot (x_n-x_m))}{2K(\kappa)}\right) \\ \nonumber
%
%+ \frac{\lambda}{2}\,b^2\operatorname{sn}^2(p\cdot x_n+\theta,k)
%
&&+\frac{\lambda}{12k^2}\,b^2
\left[
-\frac{\pi^4}{K^4(k)k}\sum_{\alpha=0}^\infty(2\alpha+1)^3
  \frac{q^{\alpha+\frac12}}{1-q^{2\alpha+1}}
  \cos\left((2\alpha+1)\frac{\pi (p\cdot (x_n-x_m))}{2K(\kappa)}\right)\right. \\ \nonumber
&&\left. +(1+k^2)\frac{\pi^2}{K^2(k)k}\sum_{\alpha=0}^\infty(2\alpha+1)
  \frac{q^{\alpha+\frac12}}{1-q^{2\alpha+1}}
  \cos\left((2\alpha+1)\frac{\pi (p\cdot (x_n-x_m))}{2K(\kappa)}\right)
\right]
=
%\delta_{nm},
0,
\eea
that we can rewrite as
\bea
&&
\frac{\pi^2}{K^2(k)k}\sum_{\alpha=0}^\infty(2\alpha+1){\hat p}^2_\alpha
  \frac{q^{\alpha+\frac12}}{1-q^{2\alpha+1}}
  \cos\left((2\alpha+1)\frac{\pi (p\cdot (x_n-x_m))}{2K(\kappa)}\right) \\ \nonumber
&&+ M^2
\frac{\pi^2}{K^2(k)k}\sum_{\alpha=0}^\infty(2\alpha+1)
  \frac{q^{\alpha+\frac12}}{1-q^{2\alpha+1}}
  \cos\left((2\alpha+1)\frac{\pi (p\cdot (x_n-x_m))}{2K(\kappa)}\right) \\ \nonumber
%
%+ \frac{\lambda}{2}\,b^2\operatorname{sn}^2(p\cdot x_n+\theta,k)
%
&&+\frac{\lambda}{12k^2}\,b^2
\left[
-\frac{\pi^4}{K^4(k)k}\sum_{\alpha=0}^\infty(2\alpha+1)^3
  \frac{q^{\alpha+\frac12}}{1-q^{2\alpha+1}}
  \cos\left((2\alpha+1)\frac{\pi (p\cdot (x_n-x_m))}{2K(\kappa)}\right)\right. \\ \nonumber
&&\left. +(1+k^2)\frac{\pi^2}{K^2(k)k}\sum_{\alpha=0}^\infty(2\alpha+1)
  \frac{q^{\alpha+\frac12}}{1-q^{2\alpha+1}}
  \cos\left((2\alpha+1)\frac{\pi (p\cdot (x_n-x_m))}{2K(\kappa)}\right)
\right]
=
%\delta_{nm},
0,
\eea

where ${\hat p}_\alpha^2=\frac{4}{a^2}\sum_\mu\sin^2\left((2\alpha+1)\frac{\pi}{4K(k)}p_\mu a\right)$. Eq.~(\ref{eq:DE}) grants the identity.

Consider now the solution
\be
G_{nm}^{(1)}=c\,\operatorname{sn}(p\cdot (x_n-x_m)+\theta,k)\operatorname{dn}(p\cdot (x_n-x_m)+\theta,k).
\ee
We observe that 
\be
(\operatorname{cn})'=-\operatorname{sn}\cdot \operatorname{dn},  \quad \operatorname{sn}^2\cdot\operatorname{sn}\cdot\operatorname{dn}=
\operatorname{sn}\cdot\operatorname{dn}-\operatorname{cn}^2\cdot\operatorname{sn}\cdot\operatorname{dn}=
\operatorname{sn}\cdot\operatorname{dn}+\frac{1}{3}(\operatorname{cn}^3)',
\quad
(\operatorname{cn}^3)'=
-\frac{1}{2k^2}(\operatorname{cn}'''+(1+k^2)\operatorname{cn}'),
\ee
and the Fourier series
\be
\label{eq:cn1}
\operatorname{cn}'(z,k)=-\frac{\pi^2}{K^2(k)k}\sum_{n=0}^\infty(2n+1)
  \frac{q^{n+\frac12}}{1+q^{2n+1}}
  \sin\left((2n+1)\frac{\pi z}{2K(\kappa)}\right),
  \ee
and
\be
\operatorname{cn}'''(z,k)=\frac{\pi^4}{K^4(k)k}\sum_{n=0}^\infty(2n+1)^3
  \frac{q^{n+\frac12}}{1+q^{2n+1}}
  \sin\left((2n+1)\frac{\pi z}{2K(\kappa)}\right).
\ee
By applying the Lam\'e operator, on gets
\bea
&&\Delta_L
\frac{\pi^2}{K^2(k)k}\sum_{\alpha=0}^\infty(2\alpha+1)
  \frac{q^{\alpha+\frac12}}{1+q^{2\alpha+1}}
  \sin\left((2\alpha+1)\frac{\pi (p\cdot (x_n-x_m))}{2K(\kappa)}\right) \\ \nonumber
&&-M^2
\frac{\pi^2}{K^2(k)k}\sum_{\alpha=0}^\infty(2\alpha+1)
  \frac{q^{\alpha+\frac12}}{1+q^{2\alpha+1}}
  \sin\left((2\alpha+1)\frac{\pi (p\cdot (x_n-x_m))}{2K(\kappa)}\right) \\ \nonumber
%
%+ \frac{\lambda}{2}\,b^2\operatorname{sn}^2(p\cdot x_n+\theta,k)
%
&&+\frac{\lambda}{2}\,b^2
\left[
-\frac{\pi^2}{K^2(k)k}\sum_{\alpha=0}^\infty(2\alpha+1)
  \frac{q^{\alpha+\frac12}}{1+q^{2\alpha+1}}
  \sin\left((2\alpha+1)\frac{\pi (p\cdot (x_n-x_m))}{2K(\kappa)}\right)\right. \nonumber \\
&&-\frac{1}{6k^2}\left(
-(1+k^2)\frac{\pi^2}{K^2(k)k}\sum_{\alpha=0}^\infty(2\alpha+1)
  \frac{q^{\alpha+\frac12}}{1+q^{2\alpha+1}}
  \sin\left((2\alpha+1)\frac{\pi z}{2K(\kappa)}\right)\right. \\ \nonumber
&&\left.\left. -\frac{\pi^4}{K^4(k)k}\sum_{\alpha=0}^\infty(2\alpha+1)^3
  \frac{q^{\alpha+\frac12}}{1+q^{2\alpha+1}}
  \sin\left((2\alpha+1)\frac{\pi z}{2K(\kappa)}\right)
\right)
\right]
=
%\delta_{nm},
0,
\eea
This is the so-called zero mode o Goldstone mode of the Lam\'e equation.

A completely similar analysis can be performed for solutions of the kind $\operatorname{cn}(z,k)\cdot\operatorname{sn}(z,k)$ and $\operatorname{cn}(z,k)\cdot\operatorname{dn}(z,k)$.

\end{document}